\documentclass[journal=jacsat,manuscript=article]{achemso}
\usepackage[version=3]{mhchem}

\usepackage{amsmath}
\usepackage{bm}

\title{Total absorption of visible light in ultra-thin weakly-absorbing semiconductor gratings}

\author{Bj\"{o}rn C. P. Sturmberg}
\email{b.sturmberg@sydney.edu.au}
\affiliation{CUDOS and IPOS, School of Physics, University of Sydney, Sydney, 2006, Australia}
\altaffiliation{Equal contribution first authors}
\affiliation{CUDOS, School of Mathematical and Physical Sciences, University of Technology Sydney, Sydney, 2007, Australia}
\author{Teck K. Chong}
\affiliation{Centre for Sustainable Energy Systems, Research School of Engineering, The Australian National University, Canberra, 2601, Australia }
\altaffiliation{Equal contribution first authors}
\author{Duk-Yong Choi}
\affiliation{Laser Physics Centre, Research School of Physics and Engineering, Australian National University, Canberra, 2601, Australia}
\author{Thomas P. White}
\affiliation{Centre for Sustainable Energy Systems, Research School of Engineering, The Australian National University, Canberra, 2601, Australia }
\author{Lindsay C. Botten}
\affiliation{National Computational Infrastructure, Australian National University, Canberra, Australia}
\altaffiliation{CUDOS, School of Mathematical and Physical Sciences, University of Technology Sydney, Sydney, 2007, Australia}
\author{Kokou B. Dossou}
\author{Christopher G. Poulton}
\affiliation{CUDOS, School of Mathematical and Physical Sciences, University of Technology Sydney, Sydney, 2007, Australia}
\author{Kylie R. Catchpole}
\affiliation{Centre for Sustainable Energy Systems, Research School of Engineering, The Australian National University, Canberra, 2601, Australia }
\author{Ross C. McPhedran}
\author{C. Martijn de Sterke}
\affiliation{CUDOS and IPOS, School of Physics, University of Sydney, Sydney, 2006, Australia}

\begin{document}
\newpage
\begin{abstract}
The perfect absorption of light in subwavelength thickness layers generally relies on exotic materials, metamaterials or thick metallic gratings. Here we demonstrate that total light absorption can be achieved in ultra-thin gratings composed of conventional materials, including relatively weakly-absorbing semiconductors, which are compatible with optoelectronic applications such as photodetectors and optical modulators. We fabricate a 41 nm thick antimony sulphide grating structure that has a measured absorptance of ${A = 99.3\%}$ at a visible wavelength of 591 nm, in excellent agreement with theory. We infer that the absorption within the grating is ${A = 98.7\%}$, with only ${A = 0.6\%}$ within the silver mirror. A planar reference sample absorbs ${A = 7.7\%}$ at this wavelength.
\end{abstract}

\maketitle

\section{Introduction}

Completely absorbing light within a layer of deeply sub-wavelength thickness, with zero reflection and zero transmission, is a challenge of both fundamental theoretical interest and of importance for practical applications such as photodetectors \cite{Watts2012}, optical switches, modulators and transducers \cite{Chong2010,Zhang2012a}.
Total light absorption (TLA) can be achieved in two ways: by adiabatically introducing a complex refractive index change over the space of many wavelengths; or by creating a critically coupled resonance.
While the first approach exhibits TLA across a broad bandwidth, it is inconsistent with the use of thin films \cite{Wang2014}. The critical coupling condition required for resonant perfect absorption is typically satisfied over a modest bandwidth, but may be achieved in structures of subwavelength thickness.

\begin{figure}[tbp]%[htbp]
\centering
{\includegraphics[width=\linewidth]{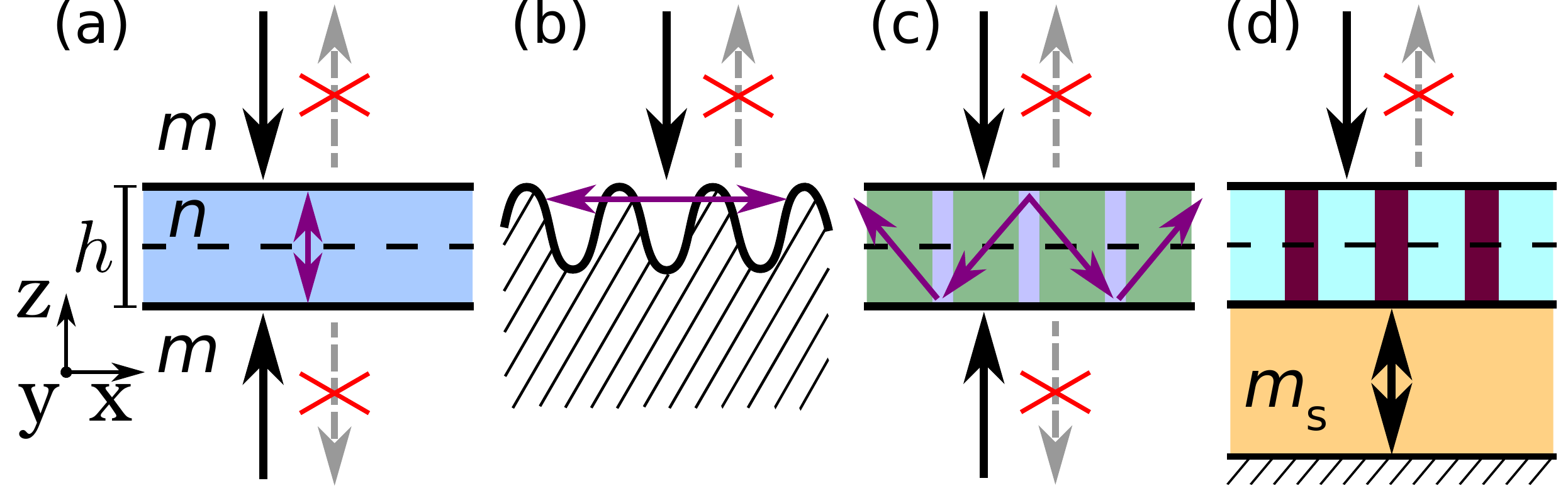}}
\caption{(a) TLA achieved by coherent illumination of a homogeneous film.
(b) TLA achieved with a surface grating on a metal, which couples light into sideways propagating SPPs.
(c) TLA in volume gratings occurs at different values of $n$ because higher order grating modes propagate with a significant sideways component.
(d) TLA in a one-port system is achieved by placing a mirror behind the absorber at a spacing that ensures the light incident from below is in phase with the light from above.}
\label{schematic}
\end{figure}

Resonant perfect absorbers typically couple light into either a longitudinal standing wave in a homogeneous (or homogenised metamaterials) layer, as in Fig.~\ref{schematic}(a) \cite{Chong2010,Zhang2012a,Salisbury1952,Tischler2006,Wan2011,Landy2008,Liu2010,Watts2012,Hagglund2010,Kats2012}, or a sideways propagating Surface Plasmon Polariton (SPP) on the surface of a corrugated metal, as in Fig.~\ref{schematic}(b) \cite{Popov2008,Maystre2013}.
TLA has also been demonstrated using plasmonic nanocomposites \cite{Hedayati2012,Hedayati2013} and plasmonic metasurfaces \cite{Zhang2014,Akselrod2015}, however plasmons are only excited by Transverse Magnetically (TM) polarized light.
The very strong absorption of unpolarized light has been achieved using crossed or bi-periodic metallic gratings \cite{Popov2008,Aydin2011}, and arrays of single layer doped graphene nano-disks \cite{Thongrattanasiri2012}.
The fabrication of metamaterial and nanoplasmonic structures is challenging because their minimum feature sizes are on the order of tens of nanometers at infrared and visible wavelengths. The use of metals also makes them incompatible with optoelectonic applications where a photocurrent must be extracted.

Here we experimentally demonstrate the total absorption of light in lamellar gratings of deeply subwavelength thickness and provide a comprehensive theoretical treatment of the phenomenon using the EMUstack package \cite{EMUstackWeb,Dossou2012,Sturmberg2015-CPC} for numerical simulations. Figures~\ref{schematic}(c) and \ref{schematic}(d) show our theoretical and experimental configurations respectively.
In our experiments we show that Transverse Electrically (TE) polarized light, where the E-field is along the grating rulings ($y$-axis in Fig.~\ref{schematic}), is totally absorbed in gratings composed of {\it relatively weakly-absorbing semiconductors} that have a complex refractive index, ${n}~=~{n'} + {\rm i}{n''}$, with ${n}'' \ll {n}'$. Such materials are abundant in nature and are compatible with optoelectronic applications. Furthermore, our structures have minimum feature sizes close to 100~nm even when targeting visible wavelengths, and can be patterned using standard techniques.
We also show numerically that TM polarized light (H-field along $y$-axis) can be totally absorbed in ultra-thin gratings, and that this requires metallic materials with ${n}''> {n}'$ (Re$(\varepsilon) < 0$), because it relies on the excitation of Surface Plasmon Polaritons (SPPs).

The paper is organized as follows: we begin by reviewing the fundamental limits to absorption in ultra-thin structure; in Sect.~\ref{deri} we derive the conditions for TLA in uniform layers; in Sect.~\ref{gratings} we investigate absorption in gratings showing that TLA occurs at very different refractive indices than in uniform layers; in Sect.~\ref{exp} we demonstrate TLA experimentally in weakly absorbing semiconductor gratings illuminated from one side; in Sect.~\ref{general} we show theoretically that TLA in ultra-thin gratings can be achieved using a very wide range of materials; and we conclude in Sect.~\ref{conclusion}.

\section{Total Light Absorption in Ultra-thin Layers}

Before investigating TLA in ultra-thin gratings we briefly examine TLA in homogeneous ultra-thin films.
An ultra-thin structure ({\it i.e.}, $|n|h \ll \lambda$, where $h$ is the layers thickness) can absorb at most 50\% of the incident power when surrounded symmetrically by uniform media (refractive index $m$) and illuminated from one side \cite{Hadley1948,Petit1989,Botten1997}.
This limit arises because the incident energy is equipartitioned between the two longitudinal modes of these structures: one has an even electric field symmetry (anti-node in the centre of the layer as in Fig.~\ref{schematic2}(a)(i)); the other has an odd symmetry (node in the centre of Fig.~\ref{schematic2}(a)(ii)) and therefore does not contribute to the absorption because it has negligible field inside the layer. The maximum absorption of ${A}~=~50\%$ occurs when the even mode is totally absorbed.
% The reflectance and transmittance of the maximally absorbing structure is $\rho = \tau = 0.25$, which are indicated with outgoing grey arrows in Fig.~\ref{schematic}(b)

In order to increase the absorption beyond 50\%, the excitation of the odd mode must be suppressed; TLA requires the structure either to not support an odd mode, or for the excitation of the odd mode to be forbidden by the symmetry of the incident field. With the odd mode not excited, TLA occurs when the even mode is fully absorbed.
In Fig.~\ref{schematic}(a) the absorbing layer is illuminated from both sides by coherent light of equal intensity, which totally suppresses the excitation of the odd mode and excites a longitudinal standing wave across the layer. This {\it coherent perfect absorption} (CPA) configuration \cite{Chong2010} produces $A=100\%$ with the same combination of $n$, $m$ and $h$ that produces ${A} = 50\%$ when illuminated from one side.

\subsection{Homogeneous (and homogenized) layers}
% \section{Analysis of Homogeneous Layers}
\label{deri}

\begin{figure}[t]
\centering
{\includegraphics[width=\linewidth]{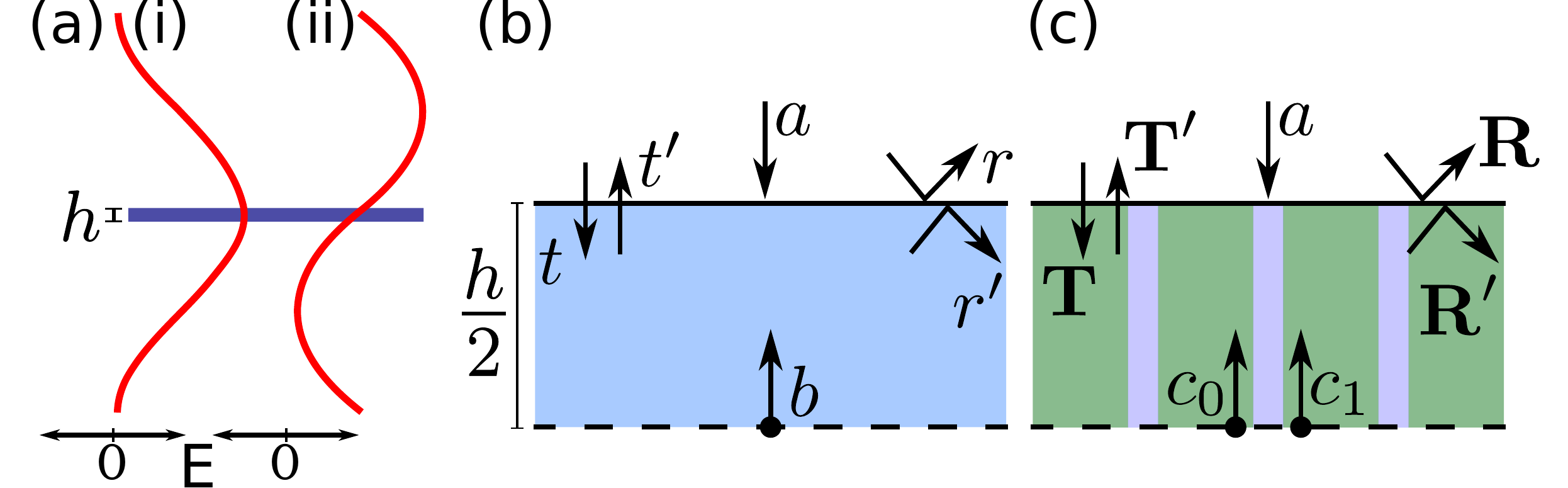}}
\caption{(a) An ultra-thin layer in a symmetric background supports an even (i), and an odd mode (ii).
The symmetry of coherent perfect absorption allows the analysis to focus on one half of the Fabry-Perot etalon: (b) a homogeneous layer; (c) a lamellar grating. Marked are the definitions of the modal amplitudes ($a$, $b$) and their reflection and transmission coefficients ($r'$, and $t'$), which in (b) are expressed in scattering matrices ($R'$, and $T'$).}
\label{schematic2}
\end{figure}

Previous studies \cite{Chong2010,Hagglund2010} have noted that TLA occurs in homogeneous layers when,
\begin{equation}
r = -\gamma^2,
\label{r_gamma}
\end{equation}
where $r = ({m - n})/({m + n})$ is the Fresnel reflection coefficient with normal incidence from the outside, and $\gamma = e^{{\rm i}{n}k_0 h/2}$ is the change in phase {\it and} amplitude acquired by a mode with complex propagation constant ${n}k_0 = 2\pi n /\lambda$ upon propagating a distance $h/2$ (see Fig.~\ref{schematic2}(b)).
While Eq.~1 has been reported previously \cite{Chong2010,Hagglund2010}, its origin has not been completely clarified.
Here we derive Eq.~1 in an intuitive, rigorous manner that generalizes to multimoded structures.

We note that Eq.~1 is consistent with critical coupling, where the loss rate of the resonance is set equal to the rate of incident energy (see Supplementary Materials for derivation). Critical coupling has been used to analyze SPP mediated TLA on corrugated metal surfaces \cite{Watts2012}. Piper {\it et al.} showed theoretically that a monolayer of graphene can achieve TLA when placed on top of a photonic crystal whose energy leakage rate matches the absorption rate of the graphene layer \cite{Piper2014}.

\subsection{Derivation of Equation~1}

In the symmetric configuration shown in Fig.~\ref{schematic}(a) we can simplify our analysis by considering the properties of one half of the Fabry-Perot etalon with one incident beam.
We begin with the resonance condition for a driven system in the presence of loss; with reference to Fig.~\ref{schematic2}(b) this is
\begin{equation}
\gamma r'\gamma b + \gamma ta = b,
% \gamma^{1/2}r'\gamma^{1/2}b + \gamma^{1/2}ta = b,
\label{lossy_res}
\end{equation}
where the first term represents a loop through the structure starting in the centre of the layer (where the amplitude $b$ is marked), and the second term describes the contribution of the driving field to the resonance (also evaluated at the centre).
In the absence of a driving field ($a=0$), Eq.~2 reduces to the Fabry-Perot resonance condition, or equivalently the condition for a bound waveguide mode.

For total absorption we require a further condition: the amplitude of the outgoing wave must vanish,
\begin{equation}
t'\gamma b + ra \equiv 0.
% t'\gamma^{1/2}b + ra \equiv 0.
\label{zero_r}
\end{equation}
This expression also consists of two terms: the leakage from the resonant mode into the superstrate; and Fresnel reflection off the top interface.
Finally, Eq.~1 is obtained by solving the simultaneous equations (Eqs.~2 and 3), using the relationship $-rr' + tt' = 1$ \cite{Born}.

\begin{figure}%[tbp]
\centering
{\includegraphics[width=0.8\linewidth]{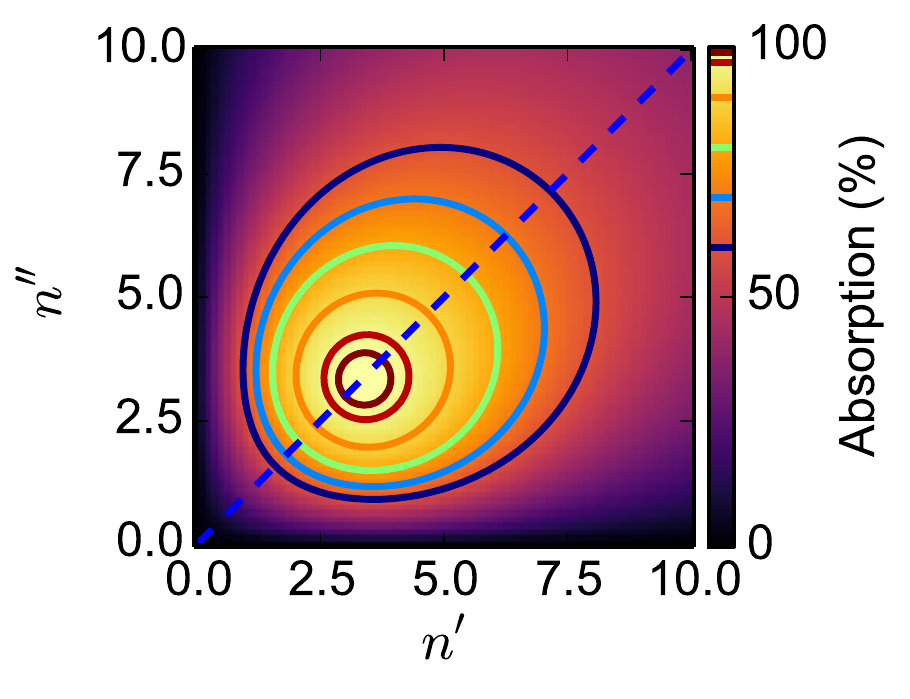}}
\caption{
Absorption of $h=\lambda/70$ thick film as a function of $n'$ and $n''$, when illuminated at normal incidence from both sides by in-phase light.}
\label{epsilon}
\end{figure}

\subsection{Requirements on \textit{\textbf{n'}} and \textit{\textbf{n''}}}
Rearranging Eq.~1 yields a transcendental expression for the complex refractive index required for TLA, which holds for structures of any thickness,
\begin{equation}
{n} = \frac{{\rm i}m}{\tan({n}k_0 h/2)}.
\label{ntan}
\end{equation}
Focussing on ultra-thin structures, we take the Taylor expansion of the tangent function in Eq.~4 for small arguments up to third order (derivation in Supplementary Materials). This yields
\begin{equation}
{n} \simeq \sqrt{\frac{{\rm i}m\lambda}{\pi h} + \frac{1}{3}{m}^2},
\label{nprime}
\end{equation}
% Equation~5 is far simpler than the expression derived by H\"{a}gglund {\it et al.} (Eq.~3 in \cite{Hagglund2010}), allowing for further analyses. The results of Eq.~5 are consistent with
which produces results consistent with the expression derived by H\"{a}gglund {\it et al.} (Eq.~3 in \cite{Hagglund2010}). While Eq.~5 does not extend to the case where the substrate and superstrate have different refractive indices, its far simpler form allows us to obtain further insights.

Equation~5 reveals that TLA in ultra-thin layers, where $\lambda/h \gg m$, is always possible in principle, and requires ${n}'~\sim~{n}''$ with ${n}''$ the slightly smaller. Furthermore $|{n}| \propto \sqrt{\lambda/h}$ in these cases, which allows TLA to be achieved with only moderately large $|n|$ even with very small $h$.
Figure~\ref{epsilon} shows the absorption as a function of the real and imaginary parts of $n$, for a uniform film of thickness $h = \lambda/70$ arranged as in Fig.~\ref{schematic}(a) and confirms that TLA occurs where ${n}'~\sim~{n}''$. Throughout our simulations we take $m = 1$.

Very few natural materials satisfy the condition ${n}'~\sim~{n}''$, making TLA a truly unusual effect; examples include dyes \cite{Tischler2006} and the phase change material VO$_2$ (when heated to precisely 342~K) \cite{Kats2012}.
TLA has also been demonstrated using metamaterials, whose subwavelength sized meta-atoms (typically a combination of inductive and capacitive metallic elements) are engineered to give an homogenized effective permittivity and permeability of ${\rm Re}(\varepsilon_{\rm eff})\approx 0$, ${\rm Re}(\mu_{\rm eff})\approx 0$ on resonance \cite{Landy2008,Liu2010,Watts2012}, which is consistent with ${n'_{\rm eff}} \sim {n''_{\rm eff}}$.
The Taylor expansion used to derive Eq.~5 is accurate only when $nk_0 h/2\ll1$, which explains how CPA has also been demonstrated in thick ($nh/\lambda > 385$) wafers of silicon at wavelengths where $n''$ is three orders of magnitude smaller than $n'$ \cite{Chong2010,Wan2011}.

\begin{figure}%[tbp]%[htbp]
\centering
{\includegraphics[width=0.8\linewidth]{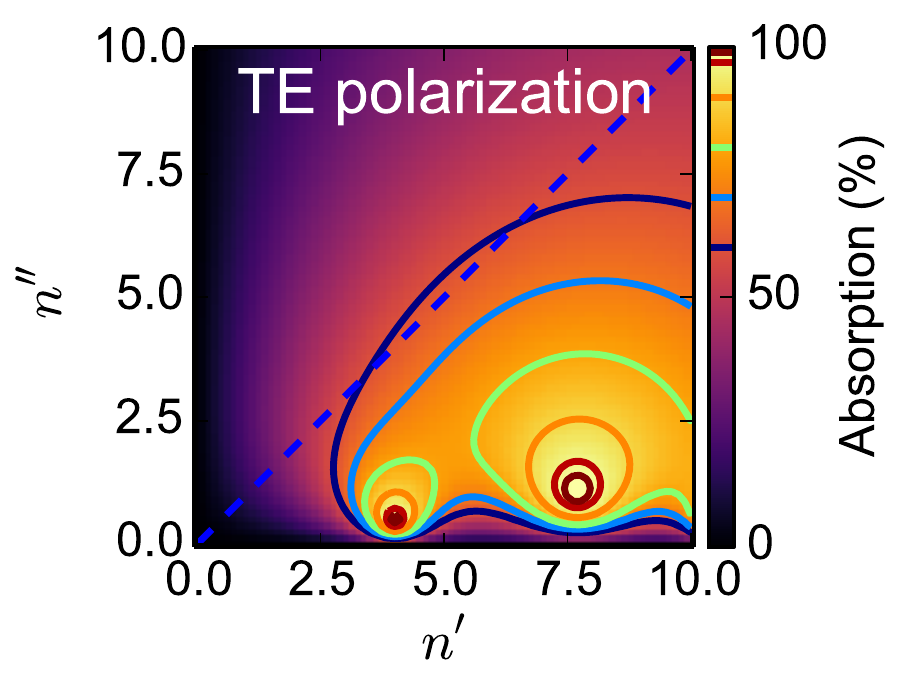}}
\caption{Absorption of $h=\lambda/70$ thick gratings as a function of $n'$ and $n''$, when illuminated at normal incidence from both sides by in-phase TE polarized light. The grating parameters are fixed at $d=66 \lambda/70$ and $f=0.5$.}
\label{epsilon-TE}
\end{figure}

\section{Total Absorption in lamellar gratings}
\label{gratings}

Having seen that TLA is difficult to achieve in uniform ultra-thin layers, we now investigate how ultra-thin gratings made of common materials can achieve TLA.
We consider the volume gratings illustrated in Figs.~\ref{schematic}(c) and \ref{schematic2}(c), a fraction $f$ of which has refractive index $n$ and the remainder of which is air, illuminated at normal incidence. The period of the gratings $d$ is chosen so that multiple Bloch modes (each corresponding to superpositions of diffraction orders) propagate within the grating, but that only the specular diffraction order propagates away from the grating in the surrounding medium. These conditions require $n_{\rm eff}d > \lambda$ and $md<\lambda$ respectively, where $n_{\rm eff}$ of the grating is calculated using the linear mixing formula of the permittivity for TE polarization (the inverse linear mixing formula must be used for TM polarization).
% ${n_{\rm eff}} = \sqrt{f_{\rm H}\epsilon_{\rm H} + (1-f_{\rm H})\epsilon_{\rm L}}$
% ${n_{\rm eff}} = 1/\sqrt{f_{\rm H}/\epsilon_{\rm H} + (1-f_{\rm H})/\epsilon_{\rm L}}$

\begin{figure}[tbp]%[htbp]
\centering
{\includegraphics[width=0.8\linewidth]{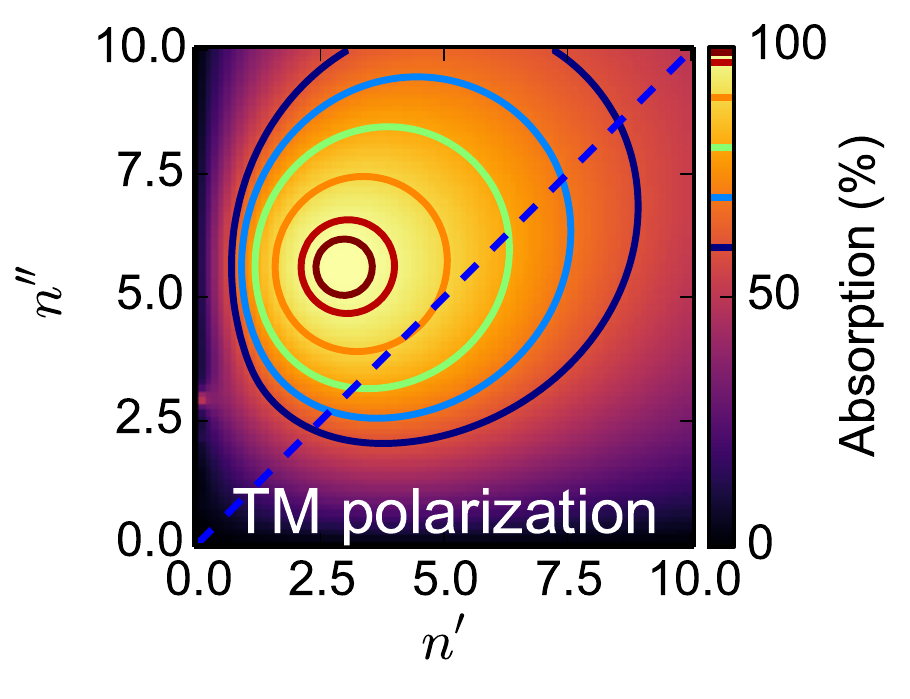}}
\caption{Absorption of $h=\lambda/70$ thick gratings as a function of $n'$ and $n''$, when illuminated at normal incidence from both sides by in-phase TM polarized light. The grating parameters are fixed at $d=66 \lambda/70$ and $f=0.5$.}
\label{epsilon-TM}
\end{figure}

The absorption of TE and TM polarized light in $h=\lambda/70$ thick lamellar gratings is shown in Figs.~\ref{epsilon-TE} and \ref{epsilon-TM} respectively, where the gratings have $d=66 \lambda/70$ and $f=0.5$. Comparing these to the results for a uniform layer of the same thickness (Fig.~\ref{epsilon}) we see that higher order diffractive grating modes drive TLA at dramatically different values of $n$ than for homogeneous layers, and that the required $n$ differs greatly between the polarizations.

\subsection{TE polarized light: slab waveguide modes}
For TE polarized light, Fig.~\ref{epsilon-TE} shows that TLA occurs with refractive indices with $n''\ll n'$ (far below the diagonal in Fig.~\ref{epsilon-TE} indicating that the gratings can be made of conventional, weakly-absorbing semiconductors.
We have established that TLA occurs due to the excitation of the fundamental TE leaky slab waveguide mode, which has no cut-off. Such guided mode resonances have previously been studied in detail for their broadband reflection properties \cite{Magnusson1992,Wang1993,Collin2014}, however TLA has not been reported using this effect.
Examining the dispersion relation of the equivalent waveguide mode of the homogenized grating \cite{Snyder1983} indicates that the absorption peak at $n'\sim 4$ corresponds to the waveguide mode being excited by the grating's first reciprocal lattice vectors, $\pm{G} = \pm2\pi/d$, whereas the peak at $n'\sim 7$ is excited by $\pm2{G}$.
While the value of $n'$ ensures that the guided mode is phase matched to the incident light, the corresponding $n''$ determines the absorption loss rate of the mode that must be equal to the mode's radiative loss rate in order to fulfil the critical coupling condition and achieve TLA.

\subsection{TM polarized light: surface plasmon polaritons}
The results for TM polarized light (using the same gratings as in Fig.~\ref{epsilon-TE} are shown in Fig.~\ref{epsilon-TM}. The refractive index that produces TLA has $n''>n'$, {\it i.e.}, Re$(\varepsilon) <0$, which implies a different underlying mechanism.
TM polarized light, unlike TE polarized light, can excite SPPs that propagate in the $x$-direction, along the interface between the surrounding medium and a metallic grating, with Re$(\varepsilon) >0$ and Re$(\varepsilon) <0$ respectively. In ultra-thin gratings, the SPPs of the top and bottom interfaces couple, producing two modes: the Long Range SPP (LRSPP); and the Short Range SPP (SRSPP), which is more lossy because it is more tightly confined within the absorber.
TLA occurs due to the SRSPP because its dominant electric field component has an even symmetry in the $xy$-plane, whereas the LRSPP has an odd symmetry. The SRSPP creates a single absorption peak in Fig.~\ref{epsilon-TM} because its propagation constant is $\beta > 0$, while Re$(\varepsilon) <0$.

\section{Experimental Demonstration}
\label{exp}

We now present our experimental demonstrations of TLA of TE polarized light using antimony sulphide ($\rm{Sb_2S_3}$) semiconductor gratings, placed above a metallic reflector as illustrated in Fig.~\ref{schematic}(d). TLA in this asymmetric configuration is driven by the same underlying physics as in Figs.~\ref{schematic}(c), but allows for TLA with only a single incident beam and is experimentally far simpler.
$\rm{Sb_2S_3}$ was chosen as the absorbing layer because it is a stable semiconductor that can be deposited in thin films using thermal evaporation and it has a suitable refractive index and absorption coefficient (in its as-deposited amorphous form) to meet the TLA requirements close to $\lambda=600~\rm{nm}$.
We emphasize that the results of Sect.~\ref{gratings} demonstrate that TLA in ultra-thin gratings is a general effect that can be achieved using a very wide range of common materials.

\begin{figure}[t]
\centering
{\includegraphics[width=0.7\linewidth]{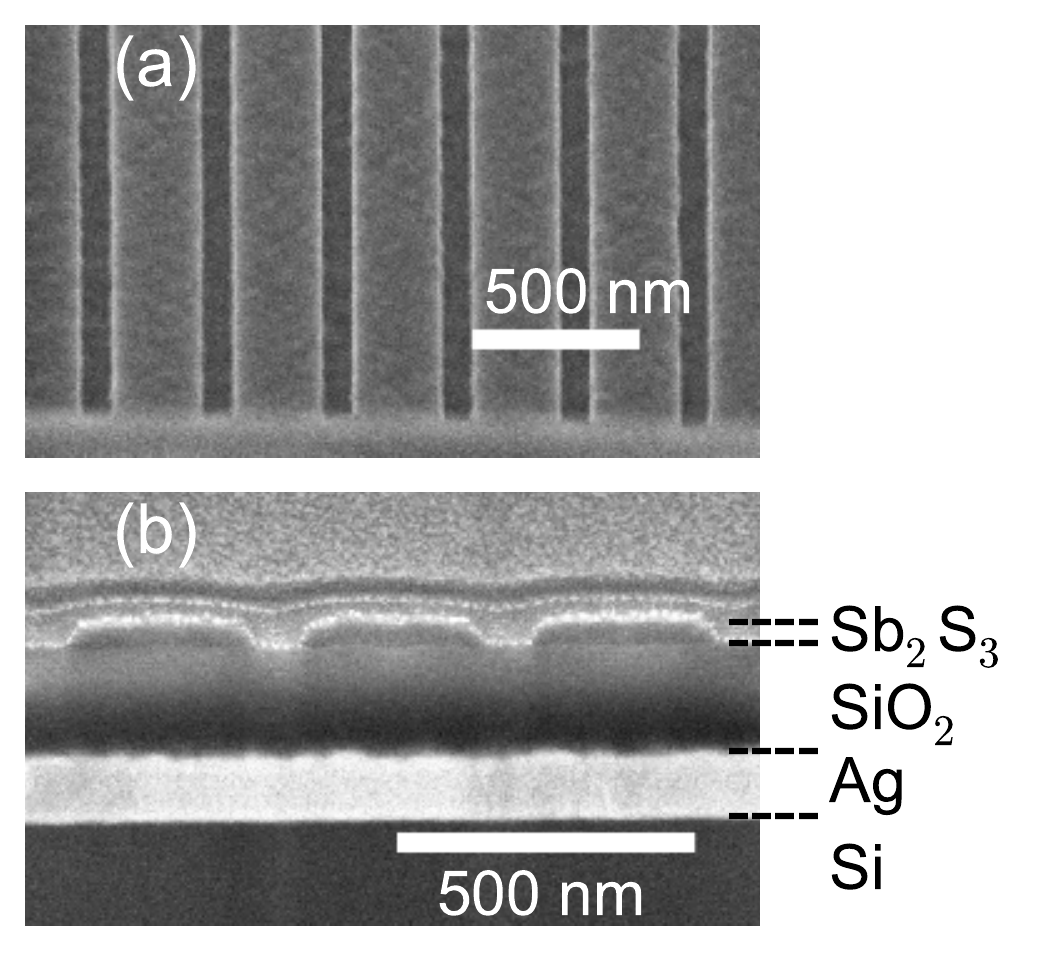}}
\caption{(a) Scanning electron micrograph at an angle of at $45^{\rm o}$ of the $\rm{Sb_2S_3}$ grating structure with $d=388$~nm etched groove width $97$~nm (designed for TLA at $\lambda=605$~nm). (b) Focused ion beam cut cross-sectional view where individual layers and the grating are clearly distinguished and labelled.
}
\label{fab}
\end{figure}

\subsection{Fabrication}
Figures~\ref{fab}(a) and \ref{fab}(b) show SEM images of a fabricated grating structure on a polished Si wafer with light incident from air. From bottom to top, the structure consists of: a $130\,\rm{nm}$ Ag reflector deposited by thermal evaporation; a ${h}_{\rm s}=245~\rm{nm}$ thick $\rm{SiO_2}$ spacer layer (refractive index ${m}_{\rm s}$) deposited by plasma-enhanced chemical-vapour-deposition (PECVD); and a $41\,{\rm nm}$ thick amorphous $\rm{Sb_2S_3}$ layer deposited by thermal evaporation. The grating was fabricated using electron-beam-lithography to define a mask in a PMMA resist, followed by an inductively-coupled plasma (ICP) etch using CHF$_3$ gas. Further details of the deposition and processing conditions are provided in the Supplementary Material.

The SiO$_2$ spacer layer is crucial for TLA in the asymmetric configuration because the light striking the absorber from below must be exactly in phase with the light incident from above to prevent the excitation of the odd mode. In the idealised case of an infinitesimally thick absorber and a perfect mirror \cite{Salisbury1952} the required spacer thickness is $h_{\rm s} = {\lambda}/{4{m}_{\rm s}}$.
However at the wavelengths of interest in our demonstration, $\lambda \sim 600$~nm, Ag is an imperfect metal, which requires $h_{\rm s} < {\lambda}/{4{m}_{\rm s}}$ (details in Supplementary Materials).

We here focus on two $41$~nm thick gratings, designed to achieve TLA at $\lambda = 591$~nm and $\lambda = 605$~nm, which have $d = 375~{\rm nm}, f = 72\%$, and $d = 385~{\rm nm}, f = 75\%$ respectively.
The measured refractive indices of ${\rm Sb_2S_3}$ at these wavelengths are $n_{\rm Sb_2S_3} = 3.342+0.096{\rm i}$ and $n_{\rm Sb_2S_3} = 3.298 + 0.074{\rm i}$ respectively, corresponding to absorption coefficients of $\alpha=2.04\times 10^4\,\rm{cm}^{-1}$ and $\alpha=1.54\times 10^4\,\rm{cm}^{-1}$, and single pass absorptances of $A=8.0\%$ and $6.1\%$.

\begin{figure}%[t]
\centering
{\includegraphics[width=\linewidth]{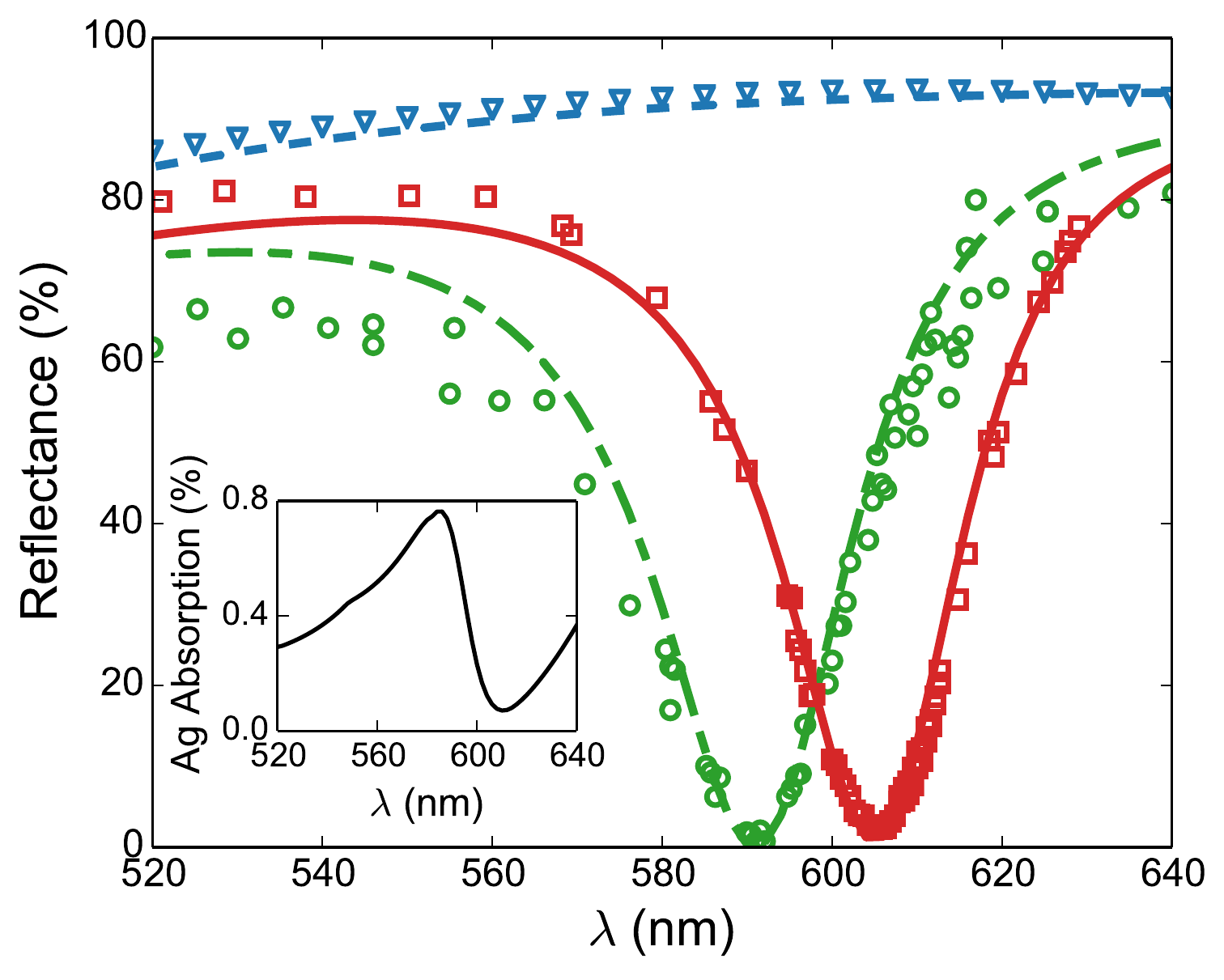}}
\caption{Reflectance of the fabricated gratings designed for TLA of TE polarized light at $\lambda = 591$~nm (green) and $\lambda = 605$~nm (red) and the planar reference structure (blue). The measured values (triangles, circles and squares respectively) and simulated predictions (dashed, dot-dashed and solid curves) show excellent agreement, in particular in the vicinity of reflectance minima. The inset shows the simulated absorption in the Ag reflector below the grating optimized for $\lambda = 591$~nm.
}
\label{measured}
\end{figure}

\subsection{Results}
Optical reflection measurements were performed in a confocal microscope using a $20\times$, NA~=~0.4 objective lens, and a broadband supercontinuum light source. Reflected light was collected by a $100\,\mu\rm{m}$ core multimode fiber and coupled into a spectrometer for detection. The reflection spectrum from a planar, unpatterned region was compared to the spectrophotometer-measured reflectance of the same sample, and this was used to calibrate the reflectance from the patterned region. The measured (symbols) and simulated (curves) reflection spectra are plotted in Fig.~\ref{measured}, and show excellent quantitative agreement, especially in the wavelengths in the vicinity of the reflectance minima, for which they are calibrated.

The minimum measured reflectance is $0.7 \pm 0.5\%$ at $591\,\rm{nm}$, which corresponds to an absorption of $A=99.3 \pm 0.5\%$ since the Ag mirror is sufficiently thick to prevent any transmission to the substrate. Although it is not possible to measure the absorption in the $\rm{Sb_2S_3}$ and in the Ag separately, we can infer these values with some confidence from simulations given the excellent agreement between the modelled and measured results. The simulations indicate that, at $\lambda=591\,\rm{nm}$, the planar $\rm{Sb_2S_3}$ film absorbs $A=7.7\%$, and the Ag mirror absorbs $A=0.3\%$ ($A=8.0\%$ total). After patterning the semiconductor layer, the calculated absorption increases to $A=98.9\%$ within the $\rm{Sb_2S_3}$ grating and $A=0.6\%$ in the Ag (the simulated absorption spectrum of the Ag layer is shown in the inset in Fig.~\ref{measured}(c)). We therefore infer $A=98.7\%$ in the experimentally realised $\rm{Sb_2S_3}$ grating.
For the second grating the maximum absorption at $\lambda=605$~nm is $A=97.4 \pm 0.5\%$, which we infer is $A=96.6\%$ in the $\rm{Sb_2S_3}$ grating. This is an even larger increase relative to the planar sample ($A=6.1\%$ at $\lambda=605$~nm) because the absorption coefficient of $\rm{Sb_2S_3}$ decreases noticeably with increasing wavelength, from $\lambda=591$~nm to $\lambda=605$~nm.

\section{Generality of TLA effect}
\label{general}

Having shown that TLA can be achieved experimentally and numerically using a specific material, we now investigate what range of material parameters are compatible with TLA in ultra-thin gratings. To do this we consider gratings with a range of $\varepsilon h /\lambda$ values, and numerically optimize the $d$ and $f$ of the grating such that the absorption is maximized.
Our optimizations were carried out in terms of $\varepsilon h /\lambda$ because we established that the properties of ultra-thin lamellar gratings depend only on this parameter. The relation $|{n}| \propto \sqrt{\lambda/h}$ therefore applies to both homogeneous layers and gratings. Our proof of this property, presented in the Supplementary Material, involves reducing the finitely conducting lamellar grating formulation \cite{Botten1981} to the grating layer formulation of Petit and Bouchitt\'{e} \cite{Bouchitte1989} in the limit of infinitesimal thickness, and is consistent with the formulation of perfectly conducting zero thickness gratings \cite{Botten1995}.

\begin{figure}%[tbp]
\centering
{\includegraphics[width=0.7\linewidth]{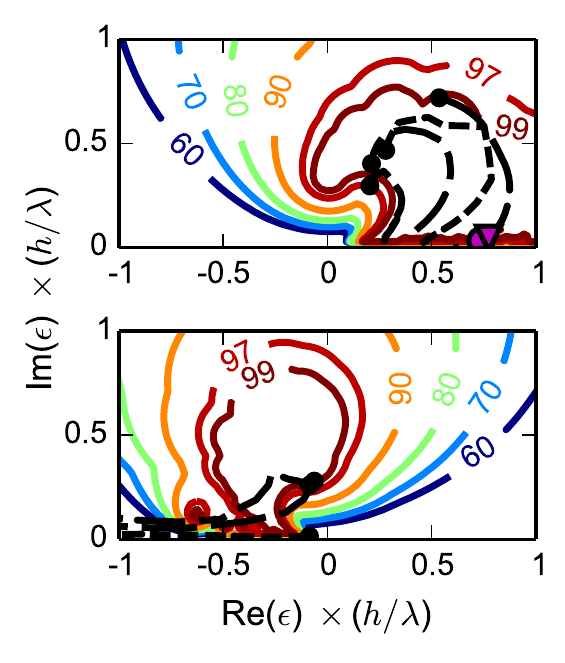}}
\caption{
Absorption as a function of Re$(\varepsilon)h/\lambda$ and Im$(\varepsilon)h/\lambda$, where $d$ and $f$ have been optimized at each $\varepsilon h/\lambda$ (coloured contours). The dashed black curves indicate $\varepsilon h/\lambda$ values of common materials at visible wavelengths between $350$~nm (marked by circle) and $800$~nm. In (a) the dashed curves correspond to $h=\lambda/30$ layers of CdTe, InP, GaAs (left to right), and $h=41 \lambda/605$ layers of Sb$_2$S$_3$ (furthest right), while in (b) the curves show the values of $h=\lambda/20$ layers of Cu, Au, and Ag (top to bottom).
The values of our experimental demonstrations in Sb$_2$S$_3$ at $\lambda=591~{\rm nm}, 605~{\rm nm}$ are marked as a magenta triangle and circle respectively.
}
\label{epsilon_b}
\end{figure}

The results of the optimization for TE and TM polarized light are shown in Figs.~\ref{epsilon_b}(a) and \ref{epsilon_b}(b) respectively. Here the coloured contours show the maximum absorption obtained at each value of $\varepsilon h /\lambda$, and the black dashed curves indicate the $\varepsilon h /\lambda$ of ultra-thin layers of common materials across the visible spectrum, $350~{\rm nm} < \lambda < 800$~nm. In Fig.~\ref{epsilon_b}(a) the dashed curves correspond to $h=\lambda/30$ layers of CdTe, InP, GaAs (left to right), and $h=41 \lambda/605$ layers of Sb$_2$S$_3$ (furthest right). The values of our experimental demonstrations in Sb$_2$S$_3$ at $\lambda=591~{\rm nm}, 605~{\rm nm}$ are marked by a magenta triangle and circle respectively.
In Fig.~\ref{epsilon_b}(b) meanwhile, the dashed curves show the values for $h/\lambda = 1/20$ thick layers of Cu, Au, and Ag \cite{Johnson1972} (top to bottom), across the same visible wavelength range.
Comparing these trajectories with the optimized absorption indicates that TLA can be achieved across almost the whole visible spectrum. The position of the dashed curves expand/contract radially from the origin when the $h/\lambda$ ratio is decreased/increased.

\subsection{Theoretical analysis of absorption in gratings}
In order to analyse TLA in gratings we examine the coupling coefficients between the grating modes and the plane waves of the surrounding medium (Fig.~\ref{schematic2}(c)); for example, $t_{00}$ is the coupling coefficient of the incident specular plane wave with the ``fundamental'' grating mode (labelled BM0), while $r'_{01}$ describes the reflection of the fundamental grating mode into the ``higher order'' grating mode (labelled BM1). We consider the case of normal incidence upon a grating in which only two modes propagate, whose amplitudes are $c_{0}$, $c_{1}$; when more modes propagate (such as at non-normal incidence) the expressions generalize with scattering matrices representing the coupling between all propagating modes.

In the case of two propagating modes, the resonance condition of BM1 is given by
\begin{equation}
\gamma_1 t_{01}a + \gamma_1 r_{01}' \gamma_0 c_{0} + \gamma_1 r_{11}' \gamma_1 c_{1} = c_{1}.
% \gamma_2^{1/2}t_{12}a + \gamma_2^{1/2} r_{12}' \gamma_1^{1/2} c_{1} + \gamma_2 r_{22}'c_{2} = c_{2}.
\label{higher_res}
\end{equation}
From left to right, these terms represent: the light incident from the surrounding medium; the coupling between BM0, and BM1; and the Fabry-Perot resonance of BM1.
Combining this with the resonance condition of BM0 (including cross coupling), and the requirement of no outgoing wave, we find that TLA occurs when
\begin{equation}
{\rm det}
\begin{bmatrix}
r_{00} & t_{00}'\gamma_0 & t_{10}'\gamma_1 \\
\gamma_0 t_{00} & \gamma_0 r_{00}'\gamma_0 -1 & \gamma_1 r_{10}' \gamma_0 \\
\gamma_1 t_{01} & \gamma_1 r_{01}' \gamma_0 & \gamma_1 r_{11}' \gamma_1 -1
% r_{11} & t_{11}'\gamma_1^{1/2} & t_{21}'\gamma_2^{1/2} \\
% \gamma_1^{1/2} t_{11} & \gamma_1 r_{11}' -1 & \gamma_2^{1/2} r_{21}' \gamma_1^{1/2} \\
% \gamma_2^{1/2}t_{12} & \gamma_2^{1/2} r_{12}' \gamma_1^{1/2} & \gamma_2r_{22}' -1
\end{bmatrix}
= 0.
\label{3_det}
\end{equation}
The top left quadrant of Eq.~7 reproduces the condition for TLA in uniform films ({\it i.e.} Eq.~2, and 3), where only one mode propagates within the absorber.
Equation~7 accurately predicts the complex refractive indices that maximise the absorption of ultra-thin gratings (Figs.~\ref{epsilon-TE} and \ref{epsilon-TM}) and we now draw on it to understand the differences between the total absorption of TE and TM polarized light.
% We now focus on effects driving TLA of TE and TM polarized light respectively, as portrayed in Figs.~\ref{epsilon-TE} and \ref{epsilon-TM}.

\subsection{Comparison between polarizations}
A striking difference between Figs.~\ref{epsilon-TE} and \ref{epsilon-TM} is that there is no absorption peak due to the excitation of the fundamental TM waveguide mode in Figs.~\ref{epsilon-TM}, even though this mode also has no cut-off.
This is because the field of this mode is concentrated almost totally within the air surrounds in ultra-thin structures, with only very little field inside the absorber, which is expressed in a modal effective index of the mode being $n_{\rm eff} \simeq 1$ \cite{footnote}.
This field distribution prevents the mode from contributing noticeably to the absorption and also means the mode is very weakly excited, because there is a negligible overlap between the guided mode and the incident field.
% This is because the effective index of the mode is $n_{\rm eff} \simeq 1$ in ultra-thin structures \cite{footnote}. This indicates that its field is concentrated almost totally within the air surrounds, with only very little field inside the absorber, preventing it from contributing noticeably to the absorption.
The analysis using Eq.~7 leads to the same conclusion; the phase of $r'_{11}$ is approximately $0$ and $\pi$, for TE and TM respectively, corresponding to a node and an anti-node of the electric field close to the grating's interface \cite{footnote2}.
In order to observe an absorption peak for TM polarized light in gratings with $n'' < n'$, the refractive index range must be increased to $|n| \approx 40$, which is unrealistic, or the thickness of a structure with $n'_{\rm eff} \sim 5$ must be increased to $h>100$~nm, at which point the structure is no longer ultra-thin.
% Having shown theoretically that TLA can be achieved in ultra-thin gratings composed of a wide variety of natural materials, we now demonstrate near total absorption of TE polarized light experimentally.

\section{Conclusion and Discussion}\
\label{conclusion}
We have shown theoretically and experimentally that ultra-thin gratings made of a wide range of weakly-absorbing semiconductors can absorb nearly 100\% of TE polarized light. We also showed theoretically that TM polarized light can be totally absorbed in ultra-thin gratings made of metals. We measure a peak absorptance of $A=99.3 \pm 0.5\%$ at $\lambda=591$~nm, in a structure with a 41~nm thick $\rm{Sb_2S_3}$ grating, where $A=98.7\%$ within the grating. Our findings show that the total absorption of shorter visible wavelengths can be straightforwardly achieved by using thinner gratings or materials with smaller $\varepsilon$.
Our gratings are far simpler to design and fabricate than existing ultra-thin perfect absorbers which rely on exotic materials and metamaterials, and may be generalized to achieve TLA of both polarizations simultaneously by using bi-periodic structures.
Ultra-thin perfect absorbers made of weakly-absorbing semiconductors may be used in optoelectronic applications such as photodetectors, where the use of semiconductors provides the possibility of extracting a photocurrent or measuring the photoresistivity.

\section{Supplementary Material}
Additional information regarding sample fabrication, optical characterization, and the effects of non-ideal metal back reflectors.
Critical coupling derivation of Eq.~1, derivation of Eq.~5 and proof of the invariance of the properties of ultra-thin gratings for constant $\varepsilon h / \lambda$.
% Simulation scripts to reproduce all numerical results using the freely available EMUstack package.

\section*{Funding Information}
This work was supported by the Australian Renewable Energy Agency, and the Australian Research Council Discovery Grant and Centre of Excellence (CE110001018) Schemes.

\section*{Acknowledgments}
We acknowledge the assistance of Y. Osario Mayon with FIB and ellipsometry measurements and X. Fu with the reflectance measurements.
Computational resources were provided by the National Computational Infrastructure, Australia and the NeCTAR Research Cloud, Australia.
Experimental facilities were provided by the Australian National Fabrication Facility and Centre for Advanced Microscopy at the Australian National University.

% \section*{Supplemental Documents}
% \emph{Optica} authors may include supplemental documents with the primary manuscript. For details, see \href{http://www.opticsinfobase.org/submit/style/supplementary-materials-optica.cfm}{Supplementary Materials in Optica}. To reference the supplementary document, the statement ``See Supplement 1 for supporting content.'' should appear at the bottom of the manuscript (above the references).

%\bigskip \noindent See \href{link}{Supplement 1} for supporting content.

% Bibliography
% \bibliography{/home/bjorn/Documents/Library/Papers/library}
% \bibliography{TLA_library}

\begin{thebibliography}{10}
\newcommand{\enquote}[1]{``#1''}

\bibitem{Watts2012}
C.~M. Watts, X.~Liu, and W.~J. Padilla, \enquote{{Metamaterial electromagnetic
  wave absorbers.}} Adv. Mater. \textbf{24}, OP98--120 (2012).

\bibitem{Chong2010}
Y.~D. Chong, L.~Ge, H.~Cao, and A.~D. Stone, \enquote{{Coherent perfect
  absorbers: Time-reversed lasers},} Phys. Rev. Lett. \textbf{105}, 1--4
  (2010).

\bibitem{Zhang2012a}
J.~Zhang, K.~F. MacDonald, and N.~I. Zheludev, \enquote{{Controlling
  light-with-light without nonlinearity},} Light Sci. Appl. \textbf{1}, 1--5
  (2012).

\bibitem{Wang2014}
K.~X. Wang, Z.~Yu, V.~Liu, M.~L. Brongersma, T.~F. Jaramillo, and S.~Fan,
  \enquote{{Nearly Total Solar Absorption in Ultrathin Nanostructured Iron
  Oxide for Efficient Photoelectrochemical Water Splitting},} ACS Photonics
  \textbf{1}, 235--240 (2014).

\bibitem{Salisbury1952}
W.~W. Salisbury, \enquote{{Absorbent Body for Electromagnetic Waves},}  (1952).

\bibitem{Tischler2006}
J.~R. Tischler, M.~S. Bradley, and V.~Bulovi\'{c}, \enquote{{Critically coupled
  resonators in vertical geometry using a planar mirror and a 5 nm thick
  absorbing film.}} Opt. Lett. \textbf{31}, 2045--2047 (2006).

\bibitem{Wan2011}
W.~Wan, Y.~Chong, L.~Ge, H.~Noh, A.~D. Stone, and H.~Cao,
  \enquote{{Time-reversed lasing and interferometric control of absorption.}}
  Science \textbf{331}, 889--892 (2011).

\bibitem{Landy2008}
N.~I. Landy, S.~Sajuyigbe, J.~J. Mock, D.~R. Smith, and W.~J. Padilla,
  \enquote{{Perfect Metamaterial Absorber},} Phys. Rev. Lett. \textbf{100},
  207402 (2008).

\bibitem{Liu2010}
X.~Liu, T.~Starr, A.~F. Starr, and W.~J. Padilla, \enquote{{Infrared Spatial
  and Frequency Selective Metamaterial with Near-Unity Absorbance},} Phys. Rev.
  Lett. \textbf{104}, 207403 (2010).

\bibitem{Hagglund2010}
C.~H\"{a}gglund, S.~P. Apell, and B.~Kasemo, \enquote{{Maximized optical
  absorption in ultrathin films and its application to plasmon-based
  two-dimensional photovoltaics.}} Nano Lett. \textbf{10}, 3135--41 (2010).

\bibitem{Kats2012}
M.~A. Kats, D.~Sharma, J.~Lin, P.~Genevet, R.~Blanchard, Z.~Yang, M.~M.
  Qazilbash, D.~N. Basov, S.~Ramanathan, and F.~Capasso, \enquote{{Ultra-thin
  perfect absorber employing a tunable phase change material},} Appl. Phys.
  Lett. \textbf{101}, 221101 (2012).

\bibitem{Popov2008}
E.~Popov, D.~Maystre, R.~C. McPhedran, M.~Nevi\`{e}re, M.~C. Hutley, and G.~H.
  Derrick, \enquote{{Total absorption of unpolarized light by crossed
  gratings.}} Opt. Express \textbf{16}, 6146--55 (2008).

\bibitem{Maystre2013}
D.~Maystre, \enquote{{Diffraction gratings: An amazing phenomenon},} Comptes
  Rendus Phys. \textbf{14}, 381--392 (2013).

\bibitem{Hedayati2012}
M.~K. Hedayati, F.~Faupel, and M.~Elbahri, \enquote{{Tunable broadband
  plasmonic perfect absorber at visible frequency},} Appl. Phys. A Mater. Sci.
  Process. \textbf{109}, 769--773 (2012).

\bibitem{Hedayati2013}
M.~K. Hedayati and M.~Elbahri, \enquote{{Perfect plasmonic absorber for visible
  frequency},}  \textbf{1}, 259--261 (2013).

\bibitem{Zhang2014}
Y.~Zhang, T.~Wei, W.~Dong, K.~Zhang, Y.~Sun, X.~Chen, and N.~Dai,
  \enquote{{Vapor-deposited amorphous metamaterials as visible near-perfect
  absorbers with random non-prefabricated metal nanoparticles.}} Sci. Rep.
  \textbf{4}, 4850 (2014).

\bibitem{Akselrod2015}
G.~M. Akselrod, J.~Huang, T.~B. Hoang, P.~T. Bowen, L.~Su, D.~R. Smith, and
  M.~H. Mikkelsen, \enquote{{Large-Area Metasurface Perfect Absorbers from
  Visible to Near-Infrared},} Adv. Mater. pp. 1--7 (2015).

\bibitem{Aydin2011}
K.~Aydin, V.~E. Ferry, R.~M. Briggs, and H.~A. Atwater, \enquote{{Broadband
  polarization-independent resonant light absorption using ultrathin plasmonic
  super absorbers},} Nat. Commun. \textbf{2}, 517 (2011).

\bibitem{Thongrattanasiri2012}
S.~Thongrattanasiri, F.~H.~L. Koppens, and F.~J. {Garc\'{\i}a de Abajo},
  \enquote{{Complete Optical Absorption in Periodically Patterned Graphene},}
  Phys. Rev. Lett. \textbf{108}, 047401 (2012).

\bibitem{EMUstackWeb}
\enquote{{EMUstack: An open-source package for Bloch mode based calculations of
  scattering matrices},} .

\bibitem{Dossou2012}
K.~B. Dossou, L.~C. Botten, A.~A. Asatryan, B.~C.~P. Sturmberg, M.~A. Byrne,
  C.~G. Poulton, R.~C. McPhedran, and C.~M. de~Sterke, \enquote{{Modal
  formulation for diffraction by absorbing photonic crystal slabs},} J. Opt.
  Soc. Am. A \textbf{29}, 817--831 (2012).

\bibitem{Sturmberg2015-CPC}
B.~C.~P. Sturmberg, K.~B. Dossou, F.~J. Lawrence, C.~G. Poulton, R.~C.
  McPhedran, C.~M. de~Sterke, and L.~C. Botten, \enquote{{EMUstack: an open
  source route to insightful electromagnetic computation via the Bloch mode
  scattering matrix method},} Computer Physics Communications (In Press) .

\bibitem{Hadley1948}
L.~N. Hadley and D.~M. Dennison, \enquote{{Reflection and Transmission
  Interference Filters},} J. Opt. Soc. Am. \textbf{38}, 483 (1948).

\bibitem{Petit1989}
R.~Petit and G.~Bouchitt\'{e}, \enquote{{On The Properties Of Very Thin
  Metallic Films In Microwaves : The Concept Of An Infinitely Conducting And
  Infinitely Thin Ohmic Material Revisited},} in \enquote{Proc. SPIE 1029,
  Scatt. Diffr.},  (1989).

\bibitem{Botten1997}
L.~C. Botten, R.~C. McPhedran, N.~A. Nicorovici, and G.~H. Derrick,
  \enquote{{Periodic models for thin optimal absorbers of electromagnetic
  radiation},} Phys. Rev. B \textbf{55}, R16072--R16082 (1997).

\bibitem{Piper2014}
J.~R. Piper and S.~Fan, \enquote{{Total Absorption in a Graphene Monolayer in
  the Optical Regime by Critical Coupling with a Photonic Crystal Guided
  Resonance},} ACS Photonics \textbf{1}, 347--353 (2014).

\bibitem{Born}
M.~Born and E.~Wolf, \enquote{{Principles of Optics},}  (2005).

\bibitem{Magnusson1992}
R.~Magnusson and S.~S. Wang, \enquote{{New principle for optical filters},}
  Appl. Phys. Lett. \textbf{61}, 1022--1024 (1992).

\bibitem{Wang1993}
S.~S. Wang and R.~Magnusson, \enquote{{Theory and applications of guided-mode
  resonance filters.}} Appl. Opt. \textbf{32}, 2606--2613 (1993).

\bibitem{Collin2014}
S.~Collin, \enquote{{Nanostructure arrays in free-space: optical properties and
  applications},} Reports Prog. Phys. \textbf{77}, 126402 (2014).

\bibitem{Snyder1983}
A.~W. Snyder and J.~D. Love, \emph{{Optical Waveguide Theory}} (Chapman and
  Hall, 1983).

\bibitem{Botten1981}
L.~C. Botten, M.~Craig, R.~C. McPhedran, J.~Adams, and J.~Andrewartha,
  \enquote{{The Dielectric Lamellar Diffraction Grating},} Opt. Acta Int. J.
  Opt. \textbf{28}, 413--428 (1981).

\bibitem{Bouchitte1989}
G.~Bouchitt\'{e} and R.~Petit, \enquote{{On the concepts of a perfectly
  conducting material and of a perfectly conducting and infinitely thin
  screen},} Radio Sci. \textbf{24}, 13 (1989).

\bibitem{Botten1995}
L.~Botten, R.~McPhedran, and G.~Milton, \enquote{{Perfectly Conducting Lamellar
  Gratings: Babinet's Principle and Circuit Models},} J. Mod. Opt. \textbf{42},
  2453--2473 (1995).

\bibitem{Johnson1972}
P.~B. Johnson and R.~W. Christy, \enquote{{Optical Constants of the Noble
  Metals},} Phys. Rev. B \textbf{6}, 4370--4379 (1972).

\bibitem{footnote}
{When $|n|h/\lambda \ll 1$, the dispersion relation of TM mode remains close to
  the light line for all but very large transverse wavevectors, which are not
  excited by gratings with $d \sim \lambda$}.

\bibitem{footnote2}
{Consistent with the phases of the TE and TM Fresnel coefficients of a
  homogeneous interface at angles of incidence just beyond total internal
  reflection (equivalent transverse wavevector as BM1)}.

\end{thebibliography}

%Manual citation list

\end{document}